\documentclass[twocolumn,showpacs,preprintnumbers]{revtex4}
\usepackage{amsfonts}
\usepackage{amsmath}
\usepackage{amssymb}
\usepackage{graphicx}
\usepackage{dcolumn}
\usepackage{bm}

\input{tcilatex}

\begin{document}

\preprint{}
\title{Generating Bragg solitons in a coherent medium}
\author{Wei Jiang}
\email{jwayne@mail.ustc.edu.cn}
\author{Qun-Feng Chen}
\author{Yong-Sheng Zhang}
\author{Guang-Can Guo}
\affiliation{Key Laboratory of Quantum Information, University of Science and Technology
of China, Hefei, 230026, P. R. China}

\begin{abstract}
In this Letter we discuss the possibility of producing Bragg solitons in an
electromagnetically induced transparency medium. We show that this coherent
medium can be engineered to be a Bragg grating with a large Kerr
nonlinearity through proper arrangements of light fields. Unlike in previous
studies, the parameters of the medium can be easily controlled through
adjusting the intensities and detunings of lasers. Thus this scheme may
provide an opportunity to study the dynamics of Bragg solitons. And doing
experiments with low power lights is possible.
\end{abstract}

\maketitle

Light propagating in a periodic medium is associated with many interesting
phenomena. A fundamental change introduced by this periodicity is a
forbidden band gap in the transmission spectrum. Light can not transmit in
this medium if its frequency falls into the band gap. However if the medium
has some nonlinearity, in some cases light can transmit in it even though
its frequency lies within the forbidden frequency band. An important example
is that a periodic medium with Kerr nonlinearity may support a kind of
solitary waves called Bragg solitons\cite{Sipe}. This kind of solitons was
firstly studied by Chen and Mills\cite{Chen}, and later by many other
researchers\cite{Mills, Joseph, Aceves, Feng, EggletonJOSAB, ContiPRE}. Some
experiments in nonlinear optical fibers were reported\cite{SipeExp, Sim}. In
these experiments very high peak intensities were required because of the
small value of the nonlinear coefficient.

On the other hand, electromagnetically induced transparency(EIT) is another
fascinating phenomenon\cite{Harris1}. An optical opaque medium is rendered
to be transparent for a probe light by a coupling field in a small frequency
window. One important property is that in the same spectral region where
there is a high degree of transmission the nonlinear response $\chi ^{(3)}$
displays constructive interference, i.e., its value at resonance could be
very large. This effect is also termed as giant Kerr effect\cite{Imamoglu}.
Another attractive application is that one can make a controllable photonic
band gap by properly arranging coupling lights\cite{Andre, Andrea, Bajcsy,
ArtoniPRL, ArtoniPRE}.

In this Letter we try to combine these two applications. We will show that
through a proper geometric configuration one can make an EIT medium to have
both a tunable photonic band gap and a large Kerr nonlinearity. We propose a
scheme to generate Bragg solitons in this medium with relatively low light
power. Furthermore, unlike in the previous studies, the parameters of the
medium can be easily controlled through adjusting the intensities and
detunings of lasers. Thus this scheme may provide an opportunity to study
the dynamics of gap solitons.

First consider an ensemble of atoms with energy diagram shown in Fig. 1. A
probe laser with frequency $\omega _{p}$ is near resonant with transitions $%
\left\vert 1\right\rangle \leftrightarrow \left\vert 3\right\rangle $ and $%
\left\vert 2\right\rangle \leftrightarrow \left\vert 4\right\rangle $ (Here
we assume $\omega _{3}-\omega _{1}\simeq \omega _{4}-\omega _{2}$). The
corresponding Rabi frequencies are $\Omega _{p}$ and $\Omega _{p^{\prime }}$
respectively. A coupling field with frequency $\omega _{c}$ and Rabi
frequency $\Omega _{c}$ is tuned to $\left\vert 2\right\rangle
\leftrightarrow \left\vert 3\right\rangle $ transition. And a controlling
field with frequency $\omega _{s}$ is tuned to $\left\vert 2\right\rangle
\leftrightarrow \left\vert 5\right\rangle $ transition.%
\begin{equation*}
Fig.1
\end{equation*}

The nonlinear coefficient of the system can be obtained by calculating the
coupled amplitude equations in the perturbative regime\cite{Harris}. The
susceptibility at the probe frequency is, 
\begin{widetext}
\begin{equation}
\chi (\omega _{p})=\frac{K_{0}}{2}\frac{-4\tilde{\delta}\tilde{\Delta}_{42}%
\tilde{\Delta}_{52}+\tilde{\Delta}_{42}\left\vert \Omega _{s}\right\vert
^{2}+\tilde{\Delta}_{52}\left\vert \Omega _{p^{\prime }}\right\vert ^{2}}{4%
\tilde{\delta}\tilde{\Delta}\tilde{\Delta}_{42}\tilde{\Delta}_{52}-\tilde{%
\Delta}\tilde{\Delta}_{52}\left\vert \Omega _{p^{\prime }}\right\vert ^{2}-%
\tilde{\Delta}\tilde{\Delta}_{42}\left\vert \Omega _{s}\right\vert ^{2}-%
\tilde{\Delta}_{42}\tilde{\Delta}_{52}\left\vert \Omega _{c}\right\vert ^{2}}%
,
\end{equation}%
\end{widetext}where$\Delta _{1}=\omega _{p}-\omega _{31}$, $\Delta
_{2}=\omega _{c}-\omega _{32}$, $\Delta _{4}=\omega _{p}-\omega _{42}$, $%
\Delta _{5}=\omega _{s}-\omega _{52}$, $\tilde{\Delta}=\Delta _{1}+\frac{i}{2%
}\Gamma _{3}$, $\tilde{\delta}=$ $\Delta _{1}-\Delta _{2}+\frac{i}{2}\Gamma
_{2}$, $\tilde{\Delta}_{42}=\Delta _{1}-\Delta _{2}+\Delta _{4}+\frac{i}{2}%
\Gamma _{4}$, $\tilde{\Delta}_{52}=\Delta _{1}-\Delta _{2}+\Delta _{5}+\frac{%
i}{2}\Gamma _{5}$. $\Gamma _{2}$, $\Gamma _{3}$, $\Gamma _{4}$and $\Gamma
_{5}$ are relaxation rates for energy level 2, 3, 4 and 5 respectively. $%
K_{0}=\rho \left\vert \mu _{13}\right\vert ^{2}/\hbar \epsilon _{0}$, $\rho $
is the density of the atoms$.$ Here we assume that ($\Gamma _{3}$, $\Gamma
_{4}$, $\Gamma _{5}$, $\left\vert \Omega _{c}\right\vert $)$\gg $($%
\left\vert \Omega _{p}\right\vert $, $\left\vert \Omega _{p^{\prime
}}\right\vert $). This susceptibility has three components $\chi
^{(1)}(\omega _{p})$, $\chi ^{(3)}(\omega _{p};\omega _{s},-\omega
_{s},\omega _{p})$ and $\chi ^{(3)}(\omega _{p};\omega _{p},-\omega
_{p},\omega _{p}).$ By defining $\chi _{a}=\chi ^{(1)}(\omega _{p})+\chi
^{(3)}(\omega _{p};\omega _{s},-\omega _{s},\omega _{p})\left\vert
E_{s}\right\vert ^{2}$ and noting $\left\vert \Omega _{p^{\prime
}}\right\vert ^{2}<<$ $\left\vert \Omega _{c}\right\vert ^{2}$, ($\Delta ,$ $%
\delta $)$<<$($\Delta _{42},\Delta _{52}$) we have,%
\begin{equation}
\chi _{a}(\omega _{p})=\frac{K_{0}}{2}\frac{-4\tilde{\delta}\tilde{\Delta}%
_{52}+\left\vert \Omega _{s}\right\vert ^{2}}{4\tilde{\delta}\tilde{\Delta}%
\tilde{\Delta}_{52}-\tilde{\Delta}_{52}\left\vert \Omega _{c}\right\vert ^{2}%
},  \label{chi1}
\end{equation}%
\begin{equation}
\chi ^{(3)}(\omega _{p};\omega _{p},-\omega _{p},\omega _{p})=\frac{K_{1}}{2}%
\frac{1}{4\tilde{\delta}\tilde{\Delta}\tilde{\Delta}_{42}-\tilde{\Delta}%
_{42}\left\vert \Omega _{c}\right\vert ^{2}},  \label{chi1a}
\end{equation}%
where $K_{1}=\rho \left\vert \mu _{13}\right\vert ^{2}\left\vert \mu
_{24}\right\vert ^{2}/\epsilon _{0}\hbar ^{3}$. Eq. (\ref{chi1},\ref{chi1a})
governs the response of the medium to the incident probe field. An important
problem about practical concern is that absorption of the medium to probe
light must be kept small. Note when $\left\vert \Omega _{c}\right\vert
^{2}>\Gamma _{2}\Gamma _{3}$, there is transparent window near the two
photon resonance. The width of the transparent window is given by $\Delta
\omega _{trans}\approx \frac{\left\vert \Omega _{c}\right\vert ^{2}}{\Gamma
_{3}^{2}\sqrt{\rho \sigma L}}$, where $\sigma =3\lambda ^{2}/2\pi $ is the
absorption cross section of an atom and $L$ is length of the sample. If we
choose a $\omega _{p}$ inside this window, the probe field will experience a
negligible small absorption while the medium still maintains a large Kerr
nonlinearity. Fig. 2 gives the real and imaginary part of $\chi _{a}$ and $%
\chi ^{(3)}$. We can see near the two-photon resonance absorptions are quite
small, while $\func{Re}(\chi ^{(3)})$ has a relative large value.%
\begin{equation*}
Fig.2
\end{equation*}

Next we concentrate on engineering a medium, which has a periodic refraction
index and a Kerr nonlinearity. Consider geometric configuration shown in
Fig. 1(b). $\vec{E}_{p}$ and $\vec{E}_{c}$ are co-propagating probe and
coupling fields. A standing wave is formed by $\vec{E}_{sf}$ and $\vec{E}%
_{sb}$. So $\Omega _{s}$ is spatially modulated and has the form $\left\vert
\Omega _{s}(\vec{r})\right\vert ^{2}=\Omega _{1}^{2}\cos ^{2}(\vec{k}%
_{s}\cdot \vec{r})$, where $\vec{k}_{s}$ is wave vector of the controlling
field. A proper angle between $\vec{E}_{p}$ and $\vec{E}_{s}$ is chosen so
that $k_{s}\cos \phi =k_{B}\approx k_{p}$ is fulfilled. This angle is small
when the wavelengths of probe field and controlling field are close. Under
this assumption we can work with a simplified 1D model. We choose the
direction of wave propagation as \textit{z} axis. Substitute $\Omega _{s}(z)$
into (\ref{chi1}) and noting $n^{2}=1+\chi $, we have%
\begin{eqnarray}
n^{2}(z) &=&1+\chi _{a}(\omega )+\chi ^{(3)}(\omega _{p};\omega _{p},-\omega
_{p},\omega _{p})\left\vert E_{p}\right\vert ^{2},  \notag \\
&=&1+\bar{\chi}_{a}+\delta \chi \cos (2k_{B}z)+\chi ^{(3)}\left\vert
E_{p}\right\vert ^{2},
\end{eqnarray}%
where $\bar{\chi}_{a}=$ $\frac{K_{0}}{2}\frac{-4\tilde{\delta}\tilde{\Delta}%
_{52}+\Omega _{1}^{2}/2}{4\tilde{\delta}\tilde{\Delta}\tilde{\Delta}_{52}-%
\tilde{\Delta}_{52}\left\vert \Omega _{c}\right\vert ^{2}}$, $\delta \chi =%
\frac{K_{0}}{4}\frac{\Omega _{1}^{2}}{4\tilde{\delta}\tilde{\Delta}\tilde{%
\Delta}_{52}-\tilde{\Delta}_{52}\left\vert \Omega _{c}\right\vert ^{2}}$.
Thus we obtain a Bragg grating with a Kerr nonlinearity. The modulation
depth of $\chi _{a}$ is controlled by $\Omega _{1}$. And the amplitude of $%
\chi ^{(3)}(\omega _{p};\omega _{p},-\omega _{p},\omega _{p})$ can be
controlled by $\Omega _{c}$ and $\Delta _{42}.$ As a result of Floquet-Bloch
theorem the periodic refraction index should produce a band gap in the
transmission spectrum known as photonic band gap. However there are two
important differences between this Bragg grating and the traditional grating
formed in an optical fiber. The first one is that the refraction index and
modulation depth is strongly frequency dependent. The second one is the
effect of absorption should be taken into account. Generally speaking the
absorption can make the edge of the band gap blur or even vanish\cite%
{AbsorptionGap, AbsorptionGap1}. We can see this effect more clearly from
the numerical result that follows. We use the transfer matrix method to get
the reflection coefficient and the dispersion relation \cite{tMatrix}.%
\begin{equation*}
Fig.3
\end{equation*}

\bigskip Fig. 3 is the calculated reflectivity of the sample and dispersion
relation. We can see there is a band gap inside the EIT window. The width of
the band gap is much less than the traditional case because of the frequency
dependence of $\bar{\chi}_{a}$ and $\delta \chi .$ This is not surprising.
The modulation of refraction index comes from the interaction between atoms
with light fields, when the probe light is far off resonant this modulation
should disappear. Therefore strong refraction index modulation exists only
in the vicinity of $\Delta _{1}=0$ where two-photon detuning is small.
Consequently the band gap is considerably narrowed. Note the edges of the
gap are blurred due to small absorption. As a comparison we also give the
reflectivity and dispersion relation when absorption is not taken into
account. From the discussion above we can conclude that a relatively strong
coupling field is needed for the photonic band gap to survive.

Now we will consider the propagation of light with frequency near or falls
into the band gap. The propagation of probe light is governed by Maxwell
equations,%
\begin{equation}
\frac{\partial ^{2}E_{p}}{\partial z^{2}}-\frac{n^{2}(z)}{c^{2}}\frac{%
\partial ^{2}E_{p}}{\partial t^{2}}=0.  \label{propagation}
\end{equation}%
Decompose $E_{p}$ into a forward and backward wave, i.e.,%
\begin{equation}
E_{p}=A_{+}(z,t)e^{i(k_{p}z-\omega _{p}t)}+A_{-}(z,t)e^{-i(k_{p}z-\omega
_{p}t)},  \label{Ep}
\end{equation}%
where $A_{+}(z,t)$ and $E_{-}(z,t)$ are the envelopes of forward and
backward waves respectively. $k_{p}=n_{0}\omega _{p}/c$ is the wave vector.
Substitute Eq.~(\ref{Ep}) into Eq.~(\ref{propagation}). After applying
slowly varying envelopes approximation and expanding $\chi _{a}$, $\delta
\chi $ and $\chi ^{(3)}$ around $\omega _{p}$, we have,

\begin{widetext}
\begin{eqnarray}
\partial _{z}A_{+}+v_{g}^{-1}\partial _{t}A_{+}-i\kappa e^{-2i\Delta
kz}A_{-}-i\gamma (\left\vert A_{+}\right\vert ^{2}+2\left\vert
A_{-}\right\vert ^{2})A_{+} &=&0,  \label{coupledmodeEq} \\
-\partial _{z}A_{-}+v_{g}^{-1}\partial _{t}A_{-}-i\kappa e^{2i\Delta
kz}A_{+}-i\gamma (\left\vert A_{-}\right\vert ^{2}+2\left\vert
A_{+}\right\vert ^{2})A_{-} &=&0,
\end{eqnarray}
\end{widetext}
where, $\Delta k=k_{p}-k_{B}$, $v_{g}=(\frac{\bar{n}}{c}+\frac{\omega _{p}}{2%
\bar{n}c}\frac{\partial \bar{\chi}}{\partial \omega })^{-1}$, $\kappa =\frac{%
\delta \chi }{4\bar{n}c}\omega _{p}$, $\gamma =\frac{\omega _{p}}{2\bar{n}c}%
\chi ^{(3)}$. We also used $\delta \chi \gg \frac{\partial \bar{\chi}}{%
\partial \omega }\gg \frac{\partial \delta \chi }{\partial \omega }$, $\chi
^{(3)}\gg \frac{\partial \chi ^{(3)}}{\partial \omega }$ in deriving Eq.(\ref%
{coupledmodeEq}). $v_{g}$ is group velocity of the light pulse inside the
medium without the light induced grating. $\kappa $ is coupling strength
between the front propagating wave $A_{+}\,$and back propagating wave $A_{-}$%
. $\gamma $ represents the self phase modulation (SPM) and cross phase
modulation (XPM) due to Kerr nonlinearity. This coupled mode equation has
been studied extensively. These equations are related to the well known
massive Thirring model of quantum field theory. Although it is
non-integrable when the SPM\ term is non-zero, shape-persevering solitary
waves can still be obtained\cite{Aceves}. The solution is,%
\begin{eqnarray}
A_{+}(z,t) &=&a_{+}\func{sech}(\zeta -i\psi /2)e^{i\theta }, \\
A_{-}(z,t) &=&a_{-}\func{sech}(\zeta +i\psi /2)e^{i\theta },
\end{eqnarray}

where%
\begin{eqnarray*}
a_{\pm } &=&\pm (\frac{1\pm \nu }{1\mp \nu })^{1/4}\sqrt{\frac{\kappa (1-\nu
^{2})}{\gamma (2-\nu ^{2})}}\sin \psi , \\
\zeta  &=&\frac{z-V_{G}t}{\sqrt{1-\nu ^{2}}}\kappa \sin \psi , \\
\theta  &=&\frac{\nu (z-V_{G}t)}{\sqrt{1-\nu ^{2}}}\kappa \cos \psi -\frac{%
4\nu }{3-\nu ^{2}}\tan ^{-1}[\cot \frac{\psi }{2}\coth \zeta ].
\end{eqnarray*}%
This solution represents a two-parameter family of Bragg solitons. The
parameter $\nu $ is in the range $-1<\nu <1$, while the parameter $\psi $
can be chosen anywhere in the range $0<\psi <\pi $. The velocity of the
Bragg soliton is given by $V_{G}=\nu v_{g}$.

The scenario here is quite different from that in nonlinear fiber gratings.
In fibers the group velocity $v_{g}$ is always comparable with c. But in our
case there is a strong normal dispersion near the two-photon resonance
beside the dispersion induced by the grating. So $v_{g}$ can be considerably
less than c. The ultra slow group velocity will cause the width of the band
gap to decrease sharply because the band gap width $\Delta \omega \ $is
given by $2\left\vert v_{g}\kappa \right\vert $. When there is no Kerr
nonlinearity, the group velocity and the group velocity dispersion (GVD) are
given by $V_{g}=v_{g}\sqrt{1-\kappa ^{2}/\delta _{\omega }^{2}}$and $\beta
_{2}=-\frac{sgn(\delta _{\omega })\kappa ^{2}/v_{g}^{2}}{(\delta _{\omega
}^{2}-\kappa ^{2})^{3/2}}$ where $\delta _{\omega }=v_{g}^{-1}(\omega
-\omega _{p})$. Here the GVD is enhanced due to the small v$_{g}.$

Now we will discuss under what conditions would we observe the optical
solitary waves in a cold atomic clouds. The low power limit ($\gamma
P_{0}\ll \kappa $) is of particular interest because in this limit the
coupled-mode equations reduced to the nonlinear Schr\"{o}dinger equation
(NLS), where $P_{0}$ is the peak power of the pulse propagating inside the
grating. In this case the Bragg soliton is actually reduced to the
fundamental NLS solitons and is found to be stable. We take some typical
values for an $^{87}$Rb atom in our estimation. $\mu _{13}=2.5\times
10^{-29} $C$\cdot $m, $N=10^{12}$cm$^{-3}$, $\Gamma _{2}=0.01\gamma _{a}$, $%
\Gamma _{3}=\Gamma _{4}=\Gamma _{5}=\gamma _{a}$, $\Delta _{1}=\Delta _{2}=0$%
, $\Delta _{4}=5\gamma _{a}$, $\Delta _{5}=20\gamma _{a}$, $\Omega
_{c}=10\gamma _{a}$ and $\Omega _{1}=10\gamma _{a}$ where $\gamma _{a}=6$%
MHz. The results are $v_{g}\approx 4200$m/s, $\kappa \approx -2600$m$^{-1}$, 
$\gamma \approx -0.60$m/W. In order to form a band gap, $\left\vert \kappa
\right\vert L\geq 2$ is required. This can be trivially fulfilled since the
typical size of the atomic cloud is about $1\sim 5$mm corresponding to $%
\left\vert \kappa \right\vert L$ $=2.6\sim 13$. The width of the band gap is
about 0.6$\gamma _{a}$. Since the bandwidth of the input pulse should be
much less than the width of the band gap, we can also estimate the minimum
width of the input pulse $T_{FWHN}>>1/\Delta \nu =0.29\mu $s$.$ The
possibility of observing Bragg solitons depends on the soliton order $N_{s}$
and the soliton period $z_{0}$\cite{Agrawal}. These parameters are given by $%
N_{s}^{2}=\frac{(3-\nu )^{2}\gamma T_{0}^{2}v_{g}^{2}\kappa \nu ^{2}}{%
2(1-\nu ^{2})^{3/2}}P_{0}$ and $z_{0}=\frac{\pi \nu
^{2}v_{g}^{2}T_{0}^{2}\kappa }{2(1-\nu ^{2})^{3/2}}$, where $T_{0}$ is
related to the FWHM of the input pulse as $T_{FWHM}\simeq 1.76T_{0}.$ A
Bragg soliton can form only if $N_{s}>\frac{1}{2}.$ So the peak power $%
P_{in} $ required to excite the fundamental Bragg soliton can be estimated
through relation $P_{in}=\frac{2(1-\nu ^{2})^{3/2}}{\nu (3-\nu
^{2})v_{g}^{2}T_{0}^{2}\kappa \gamma }.$ Here we use $P_{in}=P_{0}\nu $.
Another important parameter is $z_{0}$ since it sets the length scale over
which optical soliton evolve. It gives the minimal length of the grating.
Note $P_{in}$ and $z_{0}$ depend only on $\nu $ and $T_{0}$ when the power
and detuning of the coupling laser is fixed. Fig. 4 gives $P_{in}(\nu )$ and 
$T_{0}(\nu )$ with different $T_{0}.$ The region where we can observe Bragg
soliton is given by $P_{in}\leq P_{c}/10\ll P_{c}$ and $z_{0}\leq L$, where $%
P_{c}$ is the power of coupling laser. $L$ is the length of the cold atomic
sample. When $T_{0}=2\mu $s the workable region of $\nu $ is $%
0.05<\left\vert \nu \right\vert <0.25$. When $T_{0}=10\mu $s, the workable
region of $\nu $ is $0.0005<\left\vert \nu \right\vert <0.05$. So we can
work in different range of parameter $\nu $ by choosing different pulse
width $T_{0}.$ The lower limit of $T_{0}$ is set by the width of the band
gap. One thing worth noting is the input power $P_{in}$ is very low.
Compared with the experiments done in nonlinear fiber gratings (typical peak
intensity required is about 10 GW/cm$^{2}$), the power requirement here is
modest. This advantage is due to the ultra slow group velocity $v_{g}$
caused by the EIT\ effect and the larger Kerr coefficient caused by the
giant Kerr effect.%
\begin{equation*}
Fig.4
\end{equation*}

In conclusion we have discussed the possibility of using a coherent medium
to produce Bragg solitons. We have shown that by using a proper geometric
configuration one can make an EIT medium to have both a tunable photonic
band gap and a large Kerr nonlinearity. This scheme has two advantages. One
is it requires very low light power. The other is it provides a large
controllability over the properties of the medium. The modulation depth of
refraction index and Kerr nonlinearity can be tuned by varying $\Omega _{1}$
and $\Omega _{c}$.

\begin{acknowledgments}
This work was funded by National Fundamental Research Program, National
Natural Science Foundation of China (Grant No. 60121503, 10674127), and
Program for NCET, the Innovation funds from Chinese Academy of Sciences.
\end{acknowledgments}

\bigskip \textbf{Figure Captions}

Fig. 1 (a) Energy diagram of the atom. (b) Geometric configuration of the
lights. Coupling beam and probe beam are co-propagating. A standing wave is
formed by a forward and a backward controlling fields $\vec{E}_{sf}$ and $%
\vec{E}_{sb}$. Small angle $\phi $ between the standing wave and probe beam
is chosen so that $k_{s}cos\phi =k_{p}$ is fulfilled.

Fig. 2 Real and imaginary parts of $\chi _{a}$ and $\chi ^{(3)}$ versus $%
\Delta _{1}/\gamma _{a}$ (see the text for parameters). Near the two-photon
resonance the absorption is small while the Kerr nonlinearity is large.

Fig. 3 (a) Reflectivity of the medium(see text for parameters). A band gap
appears near the two-photon resonance. Solid lines are results when
absorption is included while dash lines are results when there is no
absorption. (b) Dispersion relation (with absorption). d is the period of
the Bragg grating. (c) Dispersion relation (without absorption).

Fig. 4 $P_{in}(\nu )$ and $T_{0}(\nu )$ with different $T_{0}$.

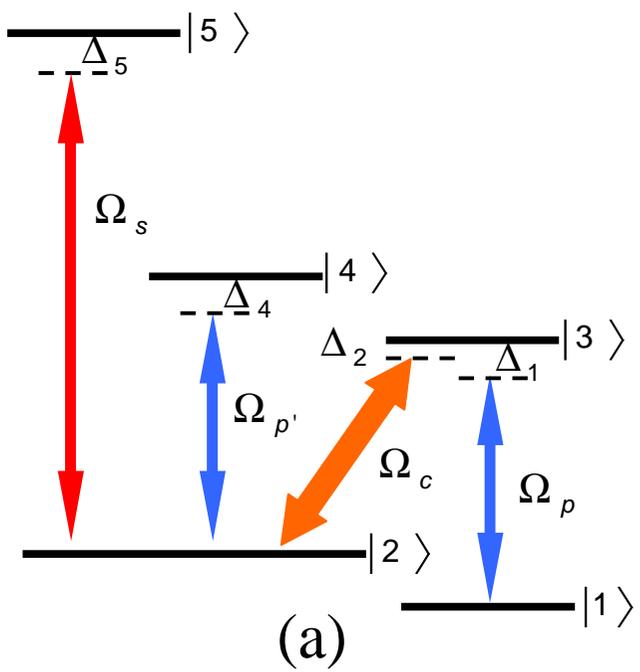
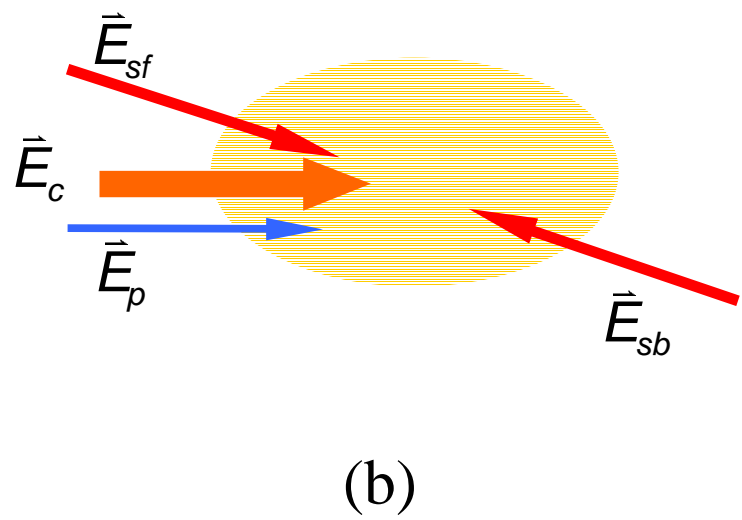
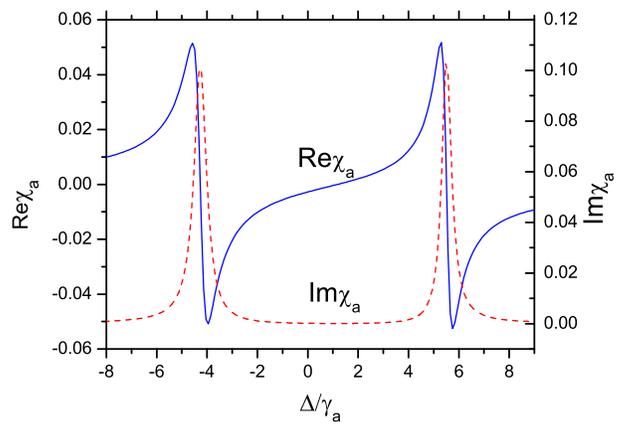
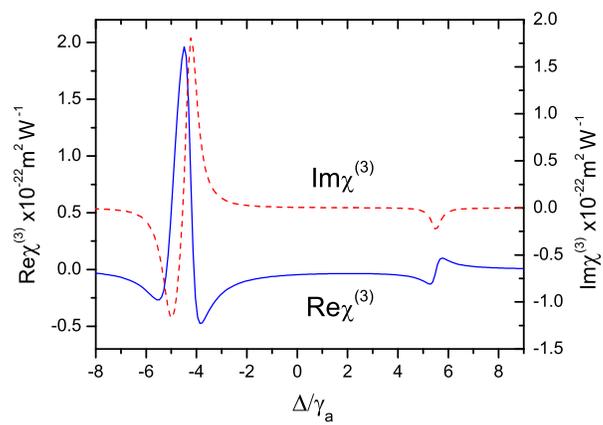
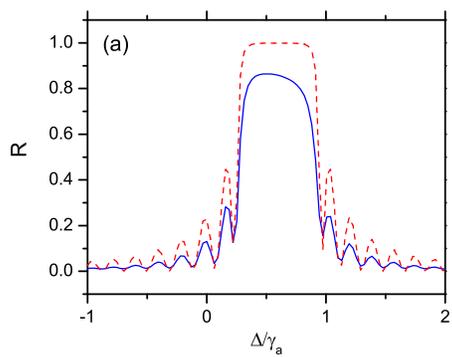
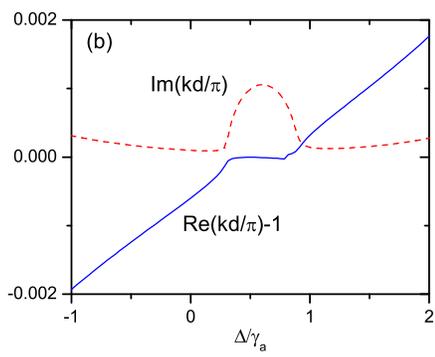
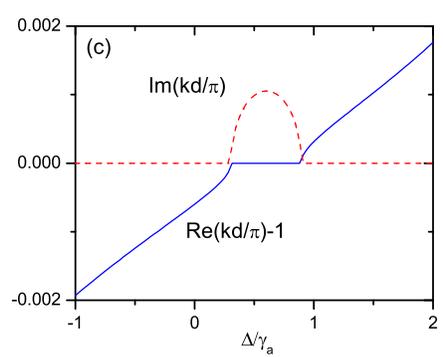
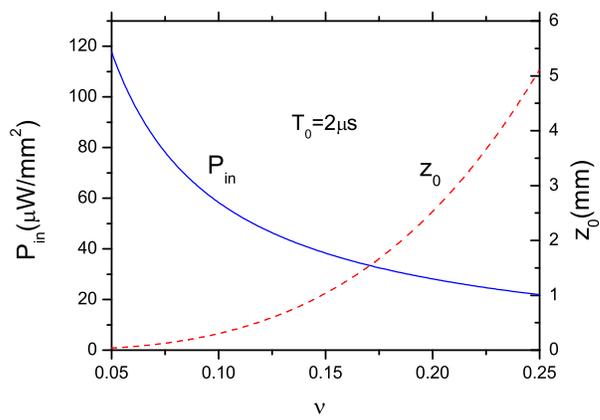
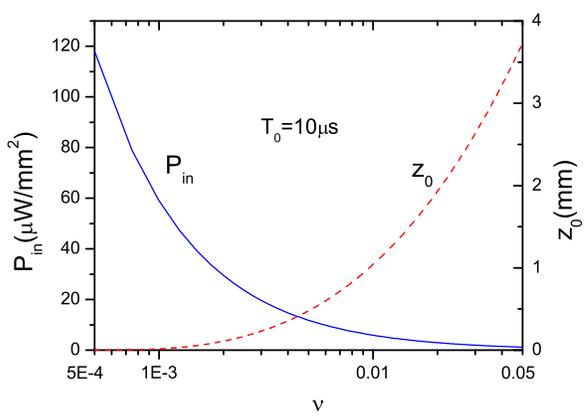
\end{document}